% bigstyle

\tolerance=10000
\magnification=1200
\raggedbottom

\baselineskip=15pt
\parskip=1\jot

\def\sk{\vskip 3\jot}

\def\heading#1{\vskip3\jot{\noindent\bf #1}}
\def\label#1{{\noindent\it #1}}
\def\QED{\hbox{\rlap{$\sqcap$}$\sqcup$}}

% Reference format

\def\ref#1;#2;#3;#4;#5.{\item{[#1]} #2,#3,{\it #4},#5.}
\def\refinbook#1;#2;#3;#4;#5;#6.{\item{[#1]} #2, #3, #4, {\it #5},#6.} 
\def\refbook#1;#2;#3;#4.{\item{[#1]} #2,{\it #3},#4.}

% Technical notations

\def\({\bigl(}
\def\){\bigr)}

% Greek alphabets

\def\ga{\gamma}
\def\de{\delta}
\def\ep{\varepsilon}

\def\la{\lambda}
\def\rh{\varrho}
\def\si{\sigma}
\def\ta{\tau}

\def\ph{\phi}

\def\ps{\psi}
\def\me{\omega}

\def\Th{\Theta}

\def\Me{\Omega}

% Calligraphic alphabet

% Boldface alphabet

\def\abs#1{\vert#1\vert}

\def\[{\big[}
\def\]{\big]}

\def\abs#1{\big\vert#1\big\vert}

{
\pageno=0
\nopagenumbers
\rightline{\tt gap.tex}
\vskip0.25in

\centerline{\bf Gap Theorems for the Delay of Circuits Simulating Finite Automata}
\vskip0.25in

\centerline{Connor Ahlbach}
\centerline{\tt Connor\_Ahlbach@hmc.edu}
\sk

\centerline{Jeremy Usatine}
\centerline{\tt Jeremy\_Usatine@hmc.edu}
\sk

\centerline{Nicholas Pippenger}
\centerline{\tt Nicholas\_Pippenger@hmc.edu}
\sk

\centerline{Department of Mathematics}
\centerline{Harvey Mudd College}
\centerline{301 Platt Boulevard}
\centerline{Claremont, CA 91711}
\vskip0.25in

\noindent{\bf Abstract:}
We study the delay (also known as depth) of circuits that simulate finite automata,
showing that only certain growth rates (as a function of the number $n$ of steps simulated) are possible.
A classic result due to Ofman (rediscovered and popularized by Ladner and Fischer) says that
delay $O(\log n)$ is always sufficient.
We show that if the automaton is ``generalized definite'', then delay $O(1)$ is sufficient, but otherwise delay $\Omega(\log n)$ is necessary; there are no intermediate growth rates.
We also consider ``physical'' (rather than ``logical'') delay, whereby we consider the lengths of wires
when inputs and outputs are laid out along a line.
In this case, delay $O(n)$ is clearly always sufficient.
We show that if the automaton is ``definite'', then delay $O(1)$ is sufficient, but otherwise delay 
$\Omega(n)$ is necessary; again there are no intermediate growth rates.
Inspired by an observation of Burks, Goldstein and von Neumann concerning the average delay due to carry propagation in ripple-carry adders, we derive conditions
for the average physical delay to be reduced from $O(n)$ to $O(\log n)$,
or to $O(1)$, 
when the inputs are independent and uniformly distributed random variables; again there are no intermediate growth rates.
Finally we consider an extension of this last result to a situation in which the inputs are not  independent and uniformly distributed, but rather are produced by a non-stationary Markov process, and in which the computation is not performed by a single automaton, but rather by a sequence of automata acting in alternating directions.
\vfill\eject
}

\heading{1. Introduction}

A classic result due to Ofman [O] (rediscovered and popularized by Ladner and Fischer [L1]) says that any computation that can be carried out by a finite automaton in $n$ steps can be performed by a circuit (a combinational logic network) with cost (also known as size, reckoned as the number of gates) $O(n)$ and  delay (also known as depth, reckoned as the maximum number of gates on any path from an input to an output) $O(\log n)$.
(In this paper, the constant factors implicit in $O(\cdots)$, $\Me(\cdots)$ and $\Th(\cdots)$ bounds
will depend on the automaton, but not of course on $n$.)
This result has been generalized by Ladner and Fischer [L1] and has, under the name ``parallel prefix computation'', become one of the central paradigms of parallel computation.
It is clear that in some cases the delay can be reduced.
If, for example, each output depends only on the first $k$ or fewer, and the most recent $k$ or fewer, inputs (for some fixed $k$), then each output depends on at most $2k$ inputs, so the delay can be reduced to $O(1)$.
We shall show in Section 2  that this case constitutes the only possible reduction.
If the automaton does not satisfy the ``generalized definite'' condition just stated, then
the delay must grow as $\Me(\log n)$, and Ofman's construction is 
(to within constant factors) the best possible.
Between $\Th(1)$ and $\Th(\log n)$, no intermediate growth rates are possible.

The results  described above  refer to the ``logical cost'' and ``logical delay'' of a circuit, which are defined in terms of numbers of gates.
These measures of complexity take no account of the cost and delay due to wires,
which in the context of integrated circuits must occupy an area proportional to their length, and which in any case must introduce a delay proportional to their length.
This circumstance raises the question of whether it is possible to obtain similar results that bound
the ``physical cost'' (where in addition to the number of gates, we charge for each wire in proportion to its length) and ``physical delay'' (where in addition to the number of gates  on a path, we charge for each wire on the path in proportion to its length).
In speaking of the lengths of wires, it is necessary to make some assumptions concerning the positions at which inputs are received and at which outputs are produced; we shall assume that the inputs and outputs for the successive steps are positioned at equidistant intervals along a line.
An obvious construction for this layout (in which ``modules'', each  simulating one step of the automaton, are
positioned at equidistant intervals along the line) yields circuits with physical cost $O(n)$ and physical delay $O(n)$.
For automata that are ``definite'' (that is, for which each output depends only on the most recent $k$ or fewer inputs, for some fixed $k$)
 the physical delay can again be reduced to $O(1)$.
We shall show in Section 3  that this case constitutes the only possible reduction.
If the automaton is not definite, then
the delay must grow as $\Me(n)$, and the obvious construction is (to within constant factors) the best possible.
Between $\Th(1)$ and $\Th(n)$, no intermediate growth rates are possible.

There are, however,  some automata for which another sort of reduction in physical delay is possible.
A typical case is given by the observation of Burks, Goldstein and von Neumann [B2] that
in a ``carry-ripple'' adder for  independent and uniformly distributed $n$-bit binary numbers,
the delay between the time the inputs are available and the time the last output is produced is
$O(\log n)$ on the average (even though it is $\Me(n)$ in the worst case), because the longest ``carry chain'' is of length $O(\log n)$ with high probability.
For a general automaton, the obvious circuit described above achieves average physical delay $O(n)$
when the input symbols are independent identically distributed random variables.
In Section 4, we shall describe conditions under which the average physical delay can be reduced to $O(\log n)$.
In fact when these conditions hold, the physical delay is $O(\log n)$ ``almost always'' (that is, with probability tending to $1$ faster than any negative power of $n$). 
Of course, if the automaton is definite, the worst-case physical delay can be reduced to $O(1)$, as described in the preceding paragraph.
We shall show that these two cases constitute the only possible reductions.
Between $\Th(1)$ and $\Th(\log n)$, and between $\Th(\log n)$ and $\Th(n)$, no intermediate growth rates are possible.

In Section 5, we shall illustrate by an example (addition of numbers
in the ``Zeckendorf representation'') how the positive result of the previous section can be extended to a situation in which the inputs are  not independently and uniformly distributed, but rather are produced by a non-stationary Markov process, and in which  the computation is not performed by a single finite automaton, but rather by a sequence of automata operating in alternating directions.
We shall show that even in this more general situation it is possible to reduce the physical delay to $O(\log n)$ almost always.

We close this introduction with the observation that our results are somewhat analogous to those
that have been obtained for the cost of circuits comprising NOT-, AND- and OR-gates with unbounded fan-in but bounded depth computing prefix-products in a semigroup.
(All gates have unit delay, but cost proportional to their number of inputs, in this model.)
The results of Yao [Y], Chandra, Fortune and Lipton [C2, C3], Dolev, Dwork, Pippenger and Wigderson  [D] and Bilardi and Preparata [B1] show that in this case as well there are just three possible growth rates, with no intermediate possibilities.
\vfill\eject

\heading{2. Logical Delay}

We use the model of finite automata due to Moore [M2].
A {\it finite automaton\/} is a sextuple $M = (A,Q,B,q_0,\de,\me)$,
where $A$ is the finite {\it input alphabet}, $Q$ is the finite set of {\it states},
$B$ is the finite {\it output alphabet}, $q_0\in Q$ is the {\it initial state},
$\de:Q\times A\to Q$ is the {\it state transition function\/} and $\me: Q\to B$ is the
{\it output function}.
We extend the state transition function $\de$ to $\de^* : Q\times A^* \to Q$ by
defining $\de^*(q, \ep) = q$ for all $q\in Q$ (where $\ep$ denotes the empty word over $A$),
and $\de^*(q,xa) = \de\(\de^*(q,x),a\)$ for all $q\in Q$, $x\in A^*$ and $a\in A$.
For brevity, we shall usually write $qx$ for $\de^*(q,x)$ when no confusion is possible.

We say that a state $q\in Q$ is {\it reachable\/} if there exists a word $x\in A^*$ such that
$q_0 x= q$.
If a state $q$ is reachable, then it is reachable by a word $x$ with length $\abs{x}$ at most $\abs{Q}$
(the cardinality of $Q$).
(The shortest word reaching $q$ cannot cause any state to be visited twice.)
We shall say that states $q_1, q_2\in Q$ are {\it distinguishable\/} if there exists a word $x\in A^*$
such that $\me(q_1 x) \not= \me(q_2 x)$.
If two states $q_1$ and $q_2$ are distinguishable, then they are distinguishable by a word $x$ with
$\abs{x}\le \abs{Q}^2$.
(The shortest word distinguishing $q_1$ and $q_2$ cannot cause any pair of states to be visited twice.)
We may restrict our attention to automata that are {\it reduced}, meaning that every state is reachable and every pair of distinct states is distinguishable (because we may always delete unreachable states and merge indistinguishable states without affecting the input-output behavior of the automaton).

The circuits we deal with are acyclic interconnections of modules by cables, where the cables carry signals drawn from various finite alphabets (such as $A$, $Q$ and $B$) and the modules produce at their output cables 
various functions (such as $\de$ and $\me$) of the signals received at their input cables.
As is well known, such circuits can be implemented as Boolean circuits, in which all cables are composed of wires that carry Boolean signals and all modules are composed of gates that compute Boolean functions of at most two arguments.
By the {\it logical cost\/} of such a circuit, we shall mean the number of gates in such a  Boolean circuit.
By the {\it logical delay\/} of such a circuit, we shall mean the maximum number of gates on any path from an input to an output
in such a Boolean circuit.
For circuits such as these, in which fan-in is bounded by two, if any output depends on $d$ different inputs, the logical delay must be at least $\log_2 d$.

We shall be concerned with circuits that {\t simulate\/} $n$ steps by a finite automaton,
receiving $n$ letters from the input alphabet at their inputs and producing $n$ letters from the output alphabet at their outputs.
The well known theorem of Ofman [O] (see also Ladner and Fischer [L1]) states that any finite automaton can be simulated by a circuit having logical cost $O(n)$ and logical delay $O(\log n)$.
We shall be concerned with the circumstances that allow the growth of the delay to be reduced.

We shall say that an automaton is {\it generalized definite\/} if each output is determined by  the $k$ earliest and the $k$ most recent inputs, for some fixed $k$.
(This notion was introduced by Ginzberg [G].)
If an automaton $M$ is generalized definite, it is clear that the logical delay can be reduced to  
$O(1)$ (because each output depends on at most $2k$ inputs). 
Our result in this section states that this situation is the only one allowing a reduction in logical delay.

\label{Theorem 2.1:}
Suppose the automaton $M$ is not generalized definite.
Then any circuit simulating $n$ steps by $M$ has logical delay $\Me(\log n)$.

\label{Proof:}
Since $M$ is not generalized definite, we can find words $xay$ and $xby$, with $a, b\in A$ and 
$x,y \in A^*$, and with $\abs{x} \ge \abs{Q} + 1$, $\abs{y}\ge \abs{Q}^2 + 1$, such that 
$\me(q_0 xay)\not=\me(q_0 xby)$.
Since the length of $x$ exceeds the number  of states of $M$, we can write $x = fg\,h$
with $f,g,h\in A^*$ and $\abs{g}\ge 1$ such that $q_0 f = q_0 fg$, and thus such that $q_0 fg^i h = q_0 x$ for all $i\ge 0$.
Since the length of $y$ exceeds the number of pairs of states of $M$, we can write $y = stu$, with $s,t,u\in A^*$ and $\abs{t}\ge 1$, such that $q_0 xa s = q_0 xa st$ and $q_0 xb s = q_0 xb st$, and thus such that
$q_0 xa st^j u = q_0 xay$ and $q_0 xb st^j u = q_0 xby$ for all $j\ge 0$.
Let $\ga = \abs{g}$ and $\ta = \abs{t}$.
Define $v = g^\ta$ and $w = t^\ga$, so that $\abs{v} = \abs{w} = \ga\ta$.
Choose $m\ge 0$.
Then the words $c_l = fg\,v^l h\, a \,s\,t\,w^{m-l} u$, for $0\le l\le m$, all have length 
$n = \rh + \ga\ta m$, where $\rh = \abs{xay} = \abs{xby}$,
and satisfy $q_0 c_l = q_0 xay$.
Similarly, the words $d_l = fg\,v^l h\, b\, s\,t\,w^{m-l} u$, for $0\le l\le m$, also all have length $n$,
and satisfy $q_0 d_l = q_0 xby$.
Since $\me(q_0 xay)\not= \me(q_0 xby)$, these $2(m+1)$ words show that the final output for these words depends on at least 
$m+1 = (n-\rh)/\ga\ta + 1 = \Me(n)$ different inputs.
Since a circuit whose output depends on $\Me(n)$ different inputs must have delay $\Me(\log n)$,
the circuits simulating any automaton that is not generalized definite must have delay $\Me(\log n)$.
\QED
\sk

\heading{3. Physical Delay}

In this section we shall consider the same model for circuits as before, but we shall measure cost and delay differently.
Specifically, in measuring the cost, we shall add to the number of gates a term proportional to the sum of the lengths of the wires used to interconnect them.
And when measuring delay, we shall add to the number of gates on a path from an input to an output a term proportional to the sum of the lengths of the wires on that path (and then take the maximum over all paths from inputs to outputs).
For this modification, we must give meaning to the notion of the ``length'' of a wire by assigning positions to input and output terminals and to gates.
In this paper we shall make what seem to be the simplest assumptions.
In a circuit that simulates $n$ steps by an automaton,
we shall allow all the terminals of wires encoding the $n$-th input, all of the terminals of wires encoding the $n$-th output, and some bounded number of gates (depending on the automaton, but not on $n$) to occupy the position of integer $n$ on the line.
(A more realistic model would insist on a layout with a minimum spacing between  input terminals, output terminals and gates,
but it clear that this would not affect cost and delay bounds  more than by constant factors, which we overlook in our analysis.)

For an upper bound, we shall consider the ``standard'' circuit in which a module that receives each input and current state, and produces each output and next state, is located at each of the positions $1, 2, \ldots, n$ along the line.
It is clear that physical cost and  delay are both $O(n)$ for this circuit.
We shall consider the circumstances under which the rate of growth of the delay can be reduced.

We shall say that an automaton is {\it definite\/} if each output is determined by  the $k$ most recent inputs, for some fixed $k$.
(This notion was introduced by Kleene [K1].)
If an automaton $M$ is  definite, it is clear that the physical delay can be reduced to  
$O(1)$ (because each output depends only on inputs  at most $k$ positions earlier). 
Our result in this section states that this situation is the only one allowing a reduction in physical delay.

\label{Theorem 3.1:}
Suppose the automaton $M$ is not definite.
Then any circuit simulating $n$ steps by $M$ has physical delay $\Me(n)$.

\label{Proof:}
Since $M$ is not definite, we can find words $xay$ and $xby$, with $a,b\in A$ and $x,y\in A^*$, and with $\abs{y}\ge \abs{Q}^2 + 1$, such that 
$\me(q_0 xay)\not=\me(q_0 xby)$.
Let $\la = \abs{xay} = \abs{xby}$.
Since the length of $y$ exceeds the number of pairs of states of $M$, we can write $y = stu$, with $s,t,u\in A^*$ and $\abs{t}\ge 1$, such that $q_0xas = q_0 xast$ and $q_0xbs = q_0xbst$, and thus such that
$q_0 xay = q_0 xast^j u$ and $q_0 xby = q_0xbst^j u$ for all $j\ge 0$.
Let  $\ta=\abs{t}$.
Choose $m\ge 0$.
The words $c_m = xast^m u$ and $d_m = xbst^m u$ have length $n = \la - \ta + \ta m$, and 
$\me(q_0 c_m) = \me(q_0 xay)\not=\me(q_0 xby) = \me(q_0 d_m)$.
Thus the final output for these words depends on an input that occurred $\ta m = \Me(n)$ positions earlier. 
But if any output depends on an input at distance $d$, the physical delay must be at least $d$.
Thus the physical delay of a circuit simulating $M$ must grow as $\Me(n)$.
\QED
\sk

\heading{4. Average Physical Delay}

Even if the automaton is not definite, so that by Theorem 3.1 there are input words that cause a physical delay $\Me(n)$, it may be the case that for some circuits, almost all input words cause a considerably smaller physical delay.
Our goal in this section is to determine conditions under which the average physical delay can be reduced to
$O(\log n)$, and to show that if these conditions are not satisfied, then average physical delay must, like the worst-case physical delay, be $\Me(n)$.
We shall also show that the average physical delay cannot be reduced below $O(\log n)$ unless the automaton is definite (in which case even the worst-case physical delay can be reduced to
$O(1)$ by the results of the preceding section).
These results will actually be shown to hold not just for the average physical delay, but ``almost always'', in a sense that will be defined more precisely below.

Our results are inspired by the observation of Burks, Goldstein and von Neumann [B2] that in a ripple-carry adder, the average length of the longest carry chain is $O(\log n)$, assuming that the inputs are independent and uniformly distributed $n$-bit binary numbers.
Further results on the average length of the longest carry chain were obtained by Claus [C4] and Knuth [K2], and
Pippenger [P] showed that the variance is $O(1)$.
These results imply that the length of the longest carry chain in a carry-ripple adder is $O(\log n)$ ``almost always''.

For our positive result in this section, we must say more about the ``standard'' circuit introduced in the previous section.
(Our negative results will of course apply to all circuits, without restriction.)
We must ensure that the module simulating the $m$-th step passes on to the module simulating the 
$(m+1)$-st step any information about the next state that can be deduced from its input signals and from any information about the previous state it receives from the module simulating the $(m-1)$-st step ``as soon as possible'' (that is, within physical delay $O(1)$ after receiving such information).
(We must also, of course, ensure that it produces its output signal as soon as possible after it can be determined from the information it has received.)
To do this, we shall assume first that the input  signals are encoded using a ``one-out-of-$\abs{A}$'' code, and that  the output signals are encoded using a ``one-out-of-$\abs{B}$'' code.
Furthermore, we shall assume the states are encoded using $2^{\abs{Q}} - 2$ Boolean signals,
one for each non-empty proper subset of the set $Q$ of states.
Each such signal will be $1$ if the state is known to belong to the given subset, and $0$ otherwise.
(We omit the signal for the empty set, which would always be $0$, and for the full set $Q$, which would always be $1$.)
Each signal produced by a given module is then a monotone Boolean function of the signals received by that module, and it can be computed with physical delay $O(1)$ by a circuit 
using the ``disjunctive normal form'' (that is, an OR of ANDs) for that Boolean function
(with AND-gates producing a $1$ as soon as $1$s are received at all their inputs, and OR-gates producing a $1$ as soon as a $1$ is received at any of their inputs).
(The conventions just described seem to be the simplest ones that allow our results to be presented.
One could also consider ``self-timed asynchronous'' circuits, such as those described by
Muller and Bartky [M3]; Mead and Conway [M1] have presented a self-timed asynchronous adder 
using that methodology.)

To simplify our results, we shall assume in this section that all automata are ``ergodic''
(that is, that their state-transition diagrams are strongly connected (so that it is possible to get from any state to any other state by some suitable input word) and that the greatest common divisor
of all cycle lengths is one (so that it is possible to reach any state by input words of any sufficiently large length)).
This entails some loss of generality  (there may be states that, once left, cannot be returned to, and there may be states that can  only be reached by input words of certain lengths), but it holds for
all the interesting examples that we know.

We shall say that $w\in A^*$ is a {\it synchronizing word\/} for the automaton $M$ and the {\it synchronized state\/} $q_1$ if
$qw = q_1$ for every state $q\in Q$.
We shall say that an automaton is {\it synchronizable\/} if it has a synchronizing word.
(These notions were introduced by Hennie [H].)
If an ergodic automaton is synchronizable, the initial state can be taken as the synchronized state,
so the synchronizing word becomes a {\it resetting word}.

We shall say that two states $q_1$ and $q_2$ are {\it mergeable\/} if there exists a word $x\in A^*$ such that $q_1 x = q_2 x$.
It is clear that if an automaton is synchronizable, every pair of states is mergeable, and \v{C}ern\'{y} [C1] has observed that the converse is also true: if every pair of states is mergeable, the automaton is synchronizable.

For simplicity, we shall assume to begin with that the successive letters of the input word are independent random variables, uniformly distributed over the input alphabet.
We shall say that a family of circuits simulating $n$ steps by an automaton has physical delay 
$O(\log n)$ {\it almost always\/} if, for every $c$, there exists a $d$ such that the probability that the physical delay exceeds $d\log n$ is at most $1/n^c$ for all sufficiently large $n$.
(Informally, this condition means that the distribution of the physical delay has an ``exponentially thin'' tail.)

\label{Theorem 4.1:}
Suppose the ergodic automaton $M$ is synchronizable.
Then there exist circuits simulating $n$ steps by $M$ with  physical delay $O(\log n)$ almost always
when the letters of the input word are independent and uniformly distributed over the input alphabet $A$.

\label{Proof:}
Suppose the ergodic automaton $M$ is synchronizable.
Let $r$ be a resetting word for $M$, and let $\rh = \abs{r}$.

Consider now any input word $y$ of length $m\ge 1$.
With probability at least $\pi = 1/\abs{A}^{\rh}$, this word will be followed by the word $r$ that brings $M$ to a state independent of $y$.
Thus if we consider any sufficiently long input word $z$, and look at its suffix of length
$\rh k$, the probability that this suffix by itself will not determine the state $q_0z$ of 
$M$ is at most $(1-\pi)^k$.
Let $c$ be any positive integer.
If we choose $k = (c+1)\log_{1/(1-\pi)} n = O(\log n)$, this probability will be at most $1/n^{c+1}$.
Let us now apply this result to each of the sufficiently long prefixes of an input word $x$ of length 
$n$.
There are at most $n$ such prefixes, and for each of them the probability that the corresponding output is not determined by its suffix of length $O(\log n)$ is at most $1/n^{c+1}$.
It follows that with probability at least $1-1/n^c$ every output is determined by the preceding $O(\log n)$ inputs.
Thus the physical delay of the standard circuit simulating $n$ steps of a synchronizable automaton is $O(\log n)$ almost always.
\QED

\label{Theorem 4.2:}
Suppose the ergodic automaton $M$ is not synchronizable.
Then any circuit simulating $n$ steps by $M$ has physical delay $\Me(n)$ almost always
when the letters of the input word are independent and uniformly distributed over the input alphabet $A$.

\label{Proof:}
Suppose the ergodic automaton $M$ is not synchronizable.
Since $M$ is ergodic, there exists a positive integer $\mu$ such that,  for any states $q, q'\in Q$,
there exists a word $v$ of length $\abs{v}=\mu$ such that $qv = q'$.
Since $M$ is not synchronizable, there exist two states $q_1, q_2\in Q$ that are not mergeable,
so that $q_1 y \not= q_2 y$ for every word $y\in A^*$.
Since $M$ is reduced, there exist, for any two distinct states $q_3, q_4\in Q$, a word
$z\in A^*$ such that $\me(q_3 z) \not= \me(q_4 z)$.
Let $\nu\ge 1$ be one more than the maximum, over all pairs of distinct states $q_3, q_4$, of the lengths of these words $z$.

Let $k\ge 1$ and $l\ge 1$ be positive integers to be chosen later, with $k = O(\log n)$ and
$l = O(\log n)$.
Let $I_1, \ldots, I_k$ be $k$ disjoint intervals of $\mu$ positions each among the first $n/3$ positions of an input word  $x$ of length $n$, and let $J_1,\ldots, J_l$ be $l$ disjoint intervals of $\nu$ positions each among the last $n/3$ positions of $x$.

Let $F_i$ denote the event ``$M$ is state either $q_1$ or $q_2$ upon exit from $I_i$''.
Each of these events has probability at least $\ph = 2/\abs{A}^\mu$ (regardless of the state of $M$ at entry to $I_1$), and the probability that none of them occur is at most $(1-\ph)^k$.
Let $c\ge 1$ be a positive integer.
Then if $k = \lceil \log_{1/(1-\ph)} (2n^c)\rceil = O(\log n)$, the probability that none of the events
$F_1, \ldots, F_k$ occur will be at most $1/2n^c$.

Suppose now that the event $F_i$ occurs.
Let $y$ denote the input subword between $I_i$ and $J_j$.
Let $q_3 = q_1 y$ and $q_4 = q_2 y$.
Let $G_j$ denote the event ``at least one of the outputs in $J_j$ depends on whether
$M$ is in state $q_3$ or state $q_4$ at entry to $J_j$''.
Each of these events has probability at least $\ps = 1/\abs{A}^\nu$ , and the probability that none of them occur is at most $(1-\ps)^l$.
Then if $l = \lceil \log_{1/(1-\ps)} (2n^c)\rceil = O(\log n)$, the probability that none of the events
$G_1, \ldots, G_k$ occur will be at most $1/2n^c$.

If at least one of the events $F_i$ occurs, and if one of the subsequent events $G_j$ occurs,
then at least one of the outputs in the last $n/3$ positions depends on at least one of the inputs in the first $n/3$ positions, and the physical delay is $\Me(n/3)$.
Thus the physical delay is $\Me(n)$ with probability at least $1 - 1/n^c$.
\QED

Theorem 4.1 applies to various schemes for multiplier recoding (see Lehman [L1, L2], Reitweisner [R] and Tocher [T] for examples), because a sufficiently long sequence of $0$s is a synchronizing word.
Another problem that can be solved with physical delay $O(\log n)$ almost always is that of carry propagation when a binary number with independent and  uniformly distributed bits is multiplied by a constant, as has been analyzed by Izsak and Pippenger [I].
On the other hand, prefix parity ($y_n = x_1 \oplus x_2 \oplus \cdots \oplus x_n$), requires average physical depth $\Me(n)$, because the two states of the reduced automaton are both reachable by words of length one, but are not mergeable.

\label{Theorem 4.3:}
Suppose the ergodic automaton $M$ is synchronizable, but not definite.
Then any circuit simulating $n$ steps by $M$ has physical delay $\Me(\log n)$ almost always
when the letters of the input word are independent and uniformly distributed over the input alphabet $A$.

\label{Proof:}
Since $M$ is ergodic and synchronizable, there exists  a resetting word $r$ for $M$. 
Since $M$ is not definite, we can find words $xay$ and $xby$, with $a,b\in A$ and $x,y\in A^*$, and with $\abs{y}\ge \abs{Q}^2$, such that $\me(q_0 xay)\not=\me(q_0 xby)$.
Since the length of $y$ exceeds the number of pairs of states of $M$, we can write $y = stu$, with $s,t,u\in A^*$ and $\abs{t}\ge 1$, such that $q_0xas = q_0 xast$ and $q_0xbs = q_0xbst$, and thus such that $q_0 xay = q_0 xast^j u$ and $q_0 xby = q_0xbst^j u$ for all $j\ge 0$.

Let $\si  = \abs{rxasu}$ and $\ta=\abs{t}$.
Then $\abs{rxast^k u} = \si + k\ta$, and the probability that $rxast^k u$ (or $rxbst^k u$) occurs
at a given position in a word is $\pi= 1/\abs{A}^{\si + k\ta}$.
If we choose $k = \lfloor (1/2\ta)\log_{\abs{A}} n\rfloor - \si$, then we have $\pi\ge 1/\sqrt{n}$
and $k\ta = \Me(\log n)$.
Thus if $rxast^k u$ (or $rxbst^k u$) occurs in a word, the physical delay will be
at least $k\ta = \Me(\log n)$.
Let $c$ be a positive integer.
In an input word of length $n$, let us choose $l = \lceil c \sqrt{n} \log n\rceil$ disjoint intervals
of length $\rh = \si + k\ta = O(\log n)$ (which we can do, because the total length of these
intervals is $O\(\sqrt{n}(\log n)^2\)$).
In each of these intervals, the subword $rxast^k u$ (or $rxbst^k u$) occurs with 
probability $\pi\ge1/\sqrt{n}$, these occurrences are independent events for the disjoint
intervals, so the probability that none of these intervals contains $rxast^k u$ (or $rxbst^k u$) 
is at most $(1-\pi)^l \le (1 - 1/\sqrt{n})^{c \sqrt{n}\log n} \le e^{-c \log n} = 1/n^{c}$.
Thus the physical delay is $\Me(\log n)$ with probability at least $1 - 1/n^c$.
\QED
\sk

\heading{5. Zeckendorf Addition}

In this section we shall apply the ideas (though not the theorems) of previous section to analyze a problem that can be solved with logarithmic physical delay almost always, even though it cannot be solved by a single finite automaton, and even though the natural probability distribution on the input words is not uniform.

The problem we analyze is that of ``Zeckendorf addition''.
The Fibonacci numbers $F_n$ for $n\ge 0$ are defined by $F_0 = 0$, $F_1 = 1$ and $F_n = 
F_{n-1} + F_{n-2}$ for $n\ge 2$.
Zeckendorf [Z] observed that any integer $M$ in the range $0\le M\le F_{n+1} - 1$ can be expressed in a unique way as a sum $M = \sum_{2\le i\le n} a_i \, F_i$ in which each $a_i$ is either $0$ or $1$
and in which no two consecutive $a_i$ are both $1$.
The word $a_n a_{n-1} \cdots a_3 a_2$ is called the {\it Zeckendorf representation\/} of $M$.
The problem  of Zeckendorf addition is to produce from the Zeckendorf representations of two integers $M$ and $N = \sum_{2\le i\le n} b_i \, F_i$ the Zeckendorf representation of their sum $M+N = \sum_{2\le i\le n+2} c_i \, F_i$ (which might require as many as two more bits than $M$ and $N$).
It is not hard to see that this problem cannot be solved by a finite automaton, but it has been shown by Frougny [F2] that it can be solved by three finite automaton that scan the input word in alternating directions, with the output words of each of the first two automata becoming the input words to their successors (see also Ahlbach, Usatine, Frougny and Pippenger [A]).
The details of these three finite automata will not concern us here.
We shall need only the following observations, which are easily seen from the descriptions
given by Ahlbach, Usatine, Frougny and Pippenger [A].
We may regard the input alphabet as $\{0,1,2\}$, where $0$ corresponds to $a_i = b_i = 0$,
$1$ corresponds to $a_i = 0$ and $b_i = 1$ or to $a_i = 1$ and $b_i = 0$, and $2$ corresponds to 
$a_i = b_i = 1$.
The first automaton is reset (synchronized into its initial state) by three consecutive $0$s in the input word, while the second and third automata are reset by two consecutive $0$s.
Furthermore, a sequence of $n\ge 3$ consecutive $0$s in the input to the first automaton results in a sequence of at least $n-2$ consecutive $0$s in its output, while a sequence of $n\ge 2$ consecutive $0$s in the input to either the second or third automaton results in a sequence of at least $n-1$ consecutive $0$s in its output.
It follows from these observations that a sequence of five consecutive $0$s in the input to the first automaton resets all the automata to their initial states at corresponding times in their operation.

If we assume that the integer $M$ is uniformly distributed in the interval $0\le M\le F_{n+1} - 1$,
then the successive bits are neither independent (because of the ``no two consecutive $1$s'' constraint) nor uniformly distributed (Filipponi and Freitag [F1] have shown that 
$\Pr[a_k = 1] = F_{k-1} \, F_{n-k+1} / F_{n+1}$.
Rather, they form a non-stationary Markov chain.
Specifically,
$$\Pr[a_2 = 1] = F_{n-1}/F_{n+1}$$
(because, among the $F_{n+1}$ $(n-1)$-bit words in question, there are $F_{n-1}$ that end with $01$).
Thus
$$\Pr[a_2 = 0] = 1 - F_{n-1}/F_{n+1} = F_{n}/F_{n+1}.$$
More generally, for $3\le k\le n$ the same reasoning yields
$$\Pr[a_k = 1 \mid a_{k-1} = 0] = F_{n-k+1}/F_{n-k+3}.$$
Thus
$$\Pr[a_k = 0 \mid a_{k-1} = 0] =  1 - F_{n-k+1}/F_{n-k+3} = F_{n-k+2}/F_{n-k+3}.$$
Of course, for $3\le k\le n$,
$$\Pr[a_k = 0 \mid a_{k-1} = 1] = 1.$$
From these results we see that the probability of $0$ in any position is at least $F_l/F_{l+1} \ge F_2/F_3 = 1/2$,
no matter what has come before.
Thus the probability of $k$ consecutive $0$s in any $k$ consecutive positions is at least $1/2^k$.
(We have described the Markov chain reading from right to left, but  a left-to-right reading yields the same chain, because the ``no two consecutive $1$s'' constraint allows a word if and only if it allows the reversal of that word.)

From these observations regarding the automata for Zeckendorf addition and the natural probability distribution on their input words, we see that in this situation again the physical delay is 
$O(\log n)$ almost always.
\sk

\heading{6. Conclusion}

Our results concerning  logical  delay appear to be definitive, but several simplifying assumptions have been made in our treatment of worst- and average-case physical delay.
Firstly, we have assumed that inputs and outputs are placed in their natural order at uniformly spaced positions along a line.
It seems likely that our negative results can be strengthened by allowing the inputs and outputs to appear in any order.
It also seems likely that these results could be generalized by allowing the inputs and outputs to be laid out (with appropriate spacing) in two or three dimensions (replacing $n$ by its square- or cube-root in the various bounds).
It should also be possible to eliminate the assumption that automata are ergodic (replacing the criterion of synchronizability by a more complicated condition).
Finally (and most ambitiously) one could try to replace the assumption of independent and uniformly distributed inputs by a more general one, say, that the inputs are generated by a stationary Markov process with finitely many states.
Such a process might make some states of the automaton effectively unreachable, or some pairs of states effectively indistinguishable, and thus will call for more far-reaching reconsideration of the problem.
\sk

\heading{7. Acknowledgment}

The research reported in this paper was supported in part by grant CCF 0917026 from the National Science Foundation.
\sk

\heading{9. References}

\ref A; C. Ahlbach, J. Usatine, Ch.\ Frougny and N. Pippenger;
``Efficient Algorithms for Zeckendorf Arithmetic'';
Fibonacci Quarterly; (to appear).

\ref B1; G. Bilardi and F. P. Preparata;
``Characterization of Associative Operations with Prefix Circuits of Constant Depth and Linear Size'';
SIAM J. Comput.; 19:2 (1990) 246--255.

\refinbook B2; A. W. Burks, H. H. Goldstein and J. von Neumann;
``Preliminary Discussion of the Logical Design of an Electronic Computing Instrument'';
in: A.~H. Taub (Ed.); Collected Works of John von Neumann;
Macmillan, 1963, v.~5, pp.~34--79. % New York

\ref C1; J. \v{C}ern\'{y};
``Pozn\'{a}mka k homog\'{e}nnym experimentom s kone\v{c}n\'{y}mi automatmi'';
Matematiko-Fyzik\'{a}lny \v{C}asopis; 14:3 (1964) 208--216.

\ref C2; A. K. Chandra, S. Fortune and R. J. Lipton;
``Unbounded Fan-In Circuits and Associative Functions'';
Proc.\ ACM Symp.\ on Theory of Computing; 15 (1983) 52--60.

\ref C3; A. K. Chandra, S. Fortune and R. J. Lipton;
``Lower Bounds for Constant Depth Circuit for Prefix Problems'';
Proc.\ Internat.\ Conf.\ on Automata, Languages and Programming; 10 (1983) 109--117.

\ref C4; V. Claus;
``Die mittlere Additionsdauer eines Paralleladdierwerks'';
Acta Informatica; 2 (1973) 283--291.

\ref D; D. Dolev, C. Dwork, N. Pippenger and A. Wigderson;
``Superconcentrators, Generalizers, and Generalized Connectors with Limited Depth'';
Proc.\ ACM Symp.\ on Theory of Computing; 15 91983) 42--51.

\refinbook F1; P. Filipponi and H. T. Freitag;
``Some Probabilistic Aspects of the Zeckendorf Decomposition of Integers'';
in: G.~E. Bergum, A.~N. Philippou and A.~F. Horadam (Ed's);
Applications of Fibonacci Numbers; Kluwer Academic Publishers, % Dordrecht
v.~7, 1998, pp.~105--114.

\ref F2; Ch.\ Frougny;
``Fibonacci Representations and Finite Automata'';
IEEE Trans.\ Inform.\ Theory; 37:2 (1991) 393--399.

\ref G; A. Ginzberg;
``About Some Properties of Definite, Reverse-Definite and Related Automata'';
IEEE Trans.\ on Electronic Computers; 15:5 (1966) 806--810.

\refbook H; F. C. Hennie;
Finite-State Models for Logical Machines;
John Wiley \& Sons, 1968. % New York

\ref I; A. Izsak and N. Pippenger;
``Carry Propagation in Multiplication by Constants'';
ACM Trans.\ Algorithms; 7:4 (2011) Art.\ 54, 11 pp..

\refinbook K1; S. C. Kleene;
``Representation of Events in Nerve Nets and Finite Automata'';
in: C.~E. Shannon and J.~McCarthy (Ed's);
Automata Studies; Princeton University Press, 1956, pp.~3--41.

\ref K2; D. E. Knuth;
``The Average Time for Carry Propagation'';
Nederl.\ Akad.\ Wettensch.\ Indag.\ Math.; 40 (1978) 238--242
 (reprinted in D.~E. Knuth, {\it Selected Papers on Analysis of Algorithms}, 
Center for the Study of Language and Information, Stanford University, 2000). % Palo Alto, CA

\ref L1; R. E. Ladner and M. J. Fischer;
``Parallel Prefix Computation'';
J. ACM; 27:4 (1980) 831-838.

\ref L2; M. Lehman;
``High-Speed Digital Multiplication'';
IRE Trans.\ Electronic Computers; 6 (1957) 204--205.

\ref L3; M. Lehman;
``Short-Cut Multiplication and Division in Automatic Binary Digital Computers'';
Proc.\ IEE; 105 B (1958) 496--504.

\refbook M1; C. Mead and L. Conway;
Introduction to VLSI Systems;
Addison-Wesley Publishing, 1980. % Reading, MA
% pp. 242--262

%
\refinbook M2; E. F. Moore;
``Gedanken-Experiments on Sequential Machines'';
in: C.~E. Shannon and J.~McCarthy (Ed's);
Automata Studies; Princeton University Press, 1956, pp.~129--153.

\item{M3} D. E. Muller and W. S. Bartky,
``A Theory of Asynchronous Circuits'',
in: {\it Proc.\ Internat.\ Symp.\ on Theory of Switching},
Harvard University Press, 1957, v.~1, pp.~204--243. % Cambridge, MA

\ref O; Yu.\ P. Ofman; 
``On the Algorithmic Complexity of Discrete Functions'';
Sov.\ Phys.\ Dokl.; 7 (1963) 589--591.

\ref P; N. Pippenger;
``Analysis of Carry Propagation in Addition: An Elementary Approach'';
J. Algorithms; 42 (2002) 317--333.

\ref R; G. W. Reitwiesner;
``Binary Arithmetic'';
Advances in Computers; 1 (1960) 232--308.

\ref T; K. D. Tocher;
``Techniques of Multiplication and Division for                                 
Automatic Binary Computers'';
Quart.\ J. Mech.\ Appl.\ Math.; 11 (1958) 364--384.

\ref Y; A. C.-C. Yao;
``Separating the Polynomial-Time Hierarchy by Oracles'';
Proc.\ IEEE Symp.\ Foundations of Computer Science; 26 (1985) 1--10.

\ref Z; \'{E}. Zeckendorf;
``Repr\'{e}sentation des nombres naturels par une somme de nombres de
Fibonacci ou de nombres de Lucas'';
Bull.\ Soc. Roy.\ Sci.\ Li\`{e}ge; 41 (1972) 179--182.

\bye